\documentclass[letterpaper, 10 pt, conference]{ieeeconf}  
\IEEEoverridecommandlockouts      
\usepackage[utf8]{inputenc}
\usepackage{amsmath}
\usepackage{amsfonts}
\usepackage{graphicx}
\usepackage{booktabs}
\usepackage{tikz}
\usetikzlibrary{positioning}
\usepackage{pgfplots}
\pgfplotsset{compat=newest}
\usepackage{xcolor}
\title{\LARGE \bf
A latent variable approach to heat load prediction in thermal grids$^\star$
}

\author{Johan Simonsson$^{1,2,\dag}$, Khalid Tourkey Atta$^{1}$, Dave Zachariah$^{3}$, Wolfgang Birk$^{1}$
\thanks{$^{1}$Control Engineering Group, Luleå University of Technology, Sweden}%
\thanks{$^{2}$Optimation AB, Uppsala, Sweden}%
\thanks{$^{3}$Department of Information Technology, Uppsala University, Sweden}%
\thanks{$^\dag$Corresponding author: johan.simonsson@ltu.se}
\thanks{$^\star$The authors want to thank Luleå Energi AB and especially Fredrik Udén for discussions and making data available. This study was financially supported by the Swedish Energy Agency under grant 43090-2, Cloudberry Datacenters.}%
}

\begin{document}

\maketitle
\thispagestyle{empty}
\pagestyle{empty}

\begin{abstract}

In this paper a new method for heat load prediction in district energy systems is proposed. The method uses a nominal model for the prediction of the outdoor temperature dependent space heating load, and a data driven latent variable model to predict the time dependent residual heat load. The residual heat load arises mainly from time dependent operation of space heating and ventilation, and domestic hot water production. The resulting model is recursively updated on the basis of a hyper-parameter free implementation that results in a parsimonious model allowing for high computational performance. The approach is applied to a single multi-dwelling building in Luleå, Sweden, predicting the heat load using a relatively small number of model parameters and easily obtained measurements. The results are compared with predictions using an artificial neural network, showing that the proposed method achieves better prediction accuracy for the validation case. Additionally, the proposed methods exhibits explainable behavior through the use of an interpretable physical model.

\end{abstract}

\section{Introduction}
District heating and cooling systems are essential technologies towards reaching climate goals and rendering an expected growth of district heating capacity in Europe \cite{Doracic2018, Connolly2012}. Moreover, 4th generation district heating networks (4GDH)\cite{Lund2014}, with lower supply temperatures generated from a wide variety of energy sources including renewable energy sources and waste heat utilization, requires novel control schemes and has ultimately increased the demand for dynamic simulation of district heating grids. In the Digital Roadmap for District Heating and Cooling \cite{DeBeaufort2018}, the \emph{digital twin}, a simulation model with embedded intelligence that is updated alongside the process, is identified as an important tool for this new generation of district heating networks.

Within this scope, accurate and computationally efficient heat load predictions for consumers are crucial. District heating grids often range over thousands of consumers, and better computationally performance means a higher resolution can be used in the grid simulation. Common use cases for a digital twin include, but are not limited to, design of new district heating grids or subnetworks, design of novel control and operation schemes, and production scheduling optimization.





Prediction of heat load for district heating consumers is not new and has been investigated by several studies, e.g using seasonal dynamic models such as SARIMAX \cite{grosswindhager_online_nodate} and by artificial neural networks (ANN) \cite{simonovic_heat_2016, idowu_applied_2016}. Models using separation of heating and domestic tap water consumption heat load has been thoroughly covered in \cite{Saint-Aubain2011}, and usage for prediction of a single house using has been covered in \cite{Bacher2013}. It should be noted that single buildings, in general, show more erratic patterns of consumption than the total consumption in the grid, and that the same level of prediction accuracy can not be reached. A comprehensive overview of predictive methods for district heating load can be found in \cite{talebi_review_2016}. 

In the present study, a heat load prediction model is proposed, using a nominal model for prediction of outdoor temperature dependent space heating load, and a latent variable model for the residual heat load, where the model is updated simultaneously and recursively. The theoretical framework is introduced in \cite{mattsson_recursive_2018} and adapted to suit the district heating heat load prediction use case. The model is verified against a single multi-dwelling building in Luleå, Sweden. 

The prediction accuracy is compared to a neural network approach, with the latent variable method showing a higher prediction accuracy for the test case. Compared to using artificial neural networks, the latent variable method allows for explainability due to the composition of the model. Here, the nominal model represents the physical building and substation, and the data driven latent variable part represents the time dependent user and control behavior. Compared to SARIMAX and more specialized models also using a split between heat load used for hot tap water consumption and space heating load, as described in \cite{Saint-Aubain2011}, the presented method produces a model that is recursively and simultaneously updated for the split model, is straight forward to aggregate for simulating larger groups of buildings, and requires less parameters.

The paper is organized as follows. The first section provide the general structure of the model and a motivation for it. The following two sections introduce the nominal model that is used to represent the thermal behavior of the building, and the latent variable model. Therein, the latent variable model for the residual heat is discussed in detail. The results from the test case are given together with the comparison with the ANN model in the Result section. The paper ends with some conclusions and outlook.


\section{General structure}
The observed heat load is the output of the substation control system including domestic hot water production, in conjunction with the thermal grid. The substation control strategy is generally not known for the entire grid, and the hot tap water use is largely stochastic in its nature. However, by using appropriate covariates we can identify the most commonly used control strategies for district heating. Here, a hybrid model approach, described in detail in \cite{mattsson_recursive_2018}, is used where $\hat{y}_{\text{nom}}(t)=\Theta \varphi(t)$ is a predictive model for the outdoor temperature dependent nominal heat load, mainly space heating. The residual load, mainly predicting time dependent space heating and ventilation, and domestic hot water production,  $\varepsilon(t) = y(t) - \hat{y}_{nom}(t)$  is then assumed to be gaussian
\begin{equation}
    \varepsilon(t) \sim \mathcal{N}(Z\gamma(t), \Sigma)
\end{equation}
with a time-varying expected value $Z \gamma(t)$, where the latent variable $Z$ and covariance matrix $\Sigma$ is unknown. The total predicted heat load is consequently
\begin{equation}
\hat{y}(t) = \hat{y}_{nom}(t) + \hat{y}_{res}(t) = \Theta \varphi(t) + Z \gamma(t).
\end{equation} 

The split model serves three purposes. For one, the nominal model is assumed to have a long term dynamic component due to possible feedback and the thermal inertia of the building, whereas we for computation efficiency do not desire to include these time lags in the user behavior-dependent part. Second, by using a suitable basis function $\gamma(t)$ it is possible to approximate both periodic and non periodic time dependent behavior without pre-processing of the data. Third, the explicit separation of outdoor temperature dependent and time dependent behavior allows for physical interpretability of the nominal model that can be exploited in future separate categorizations of the two when a larger building stock is targeted. 

Joint estimation of  the nominal model and the residual load model is based on the maximum likelihood approach with a latent variable $Z$ that is distributed as
\begin{equation}
    \text{vec}(Z) \sim \mathcal{N}(0,D).
\end{equation} 

The nominal model parameters $\Theta$ are found by the maximum likelihood approach, maximizing 
\begin{equation}
    \label{eq:p_y_omega}
    p(Y|\Omega) = \int p(Y|\Omega, Z)p(Z)dz
\end{equation}{}

\noindent where $\Omega = \{\Theta, D, \Sigma\}$. The problem may have local minima, and an estimation of the parameters is found using the Expectation Maximization technique \cite{dempster_maximum_1977} where the cost function obtained from the maximization (\ref{eq:p_y_omega}) is guaranteed to decrease monotonically. The latent variables Z can then be estimated at the optimal estimate of $\Omega$. The calculations include a data-adaptive regularizing term that produces parsimonious estimates of Z \cite{stoica_spice_2012} without user-specified hyperparameters.

Notably, the nominal model must be linear in the parameters, but nonlinear inputs can be provided if applicable. As the total heat load for the grid is predicted as the sum of distributed loads it allows for the use of specific models for consumers that cannot be predicted from readily available data, e.g. certain types of industrial plants. For the individual consumer heat load prediction, the following information is updated and stored for every time step

\begin{itemize}
\item $\Theta$ matrix – Parameters of the nominal model
\item Z matrix – Parameters of the latent variable model
\item $\Sigma$ matrix – Variance of the total model error
\item n – Number of observations used for the model
\end{itemize}

The aggregation of distributed heat loads is straight-forward since the total predicted heat load at a given time instance is the sum of individual predictions. The errors are uncorrelated, and the total predicted heat load for $N$ consumers and the total variance is given by:
\begin{equation}
\hat{y}_{\mathrm{tot}}(t) = \sum_{i=1}^{N}{\hat{y}_i(t)} \quad \text{and} \quad \Sigma_{\mathrm{tot}} = \sum_{i=1}^{N}{\Sigma_i},
\end{equation}
which can be learned recursively.

\section{Nominal model}
The space heating load is often mainly outdoor temperature dependent due to the most common control strategies, involving possible combinations of outdoor temperature dependent supply temperature from the grid operator, outdoor temperature dependent feed forward control for the substation, and indoor temperature feedback control for the substation. The latter is indirectly affected by the outdoor temperature. The control strategies can also directly or indirectly (for feedback control) depend on factors such as wind speed, solar radiation and precipitation.

The goal of the space heating control strategy is generally to keep the indoor temperature at a stable comfort temperature. However, by manipulating the stored heat in the building by raising or lowering the indoor temperature slightly, the building can be used as a passive thermal storage for the grid \cite{romanchenko_thermal_2018}. 

A simplified energy balance for the building adapted from \cite{ballarini_analysis_2012} describes the dynamics of the building
\setlength{\arraycolsep}{0.0em}
\begin{eqnarray}
    \label{eq:energybalance}
   C_{th}\frac{\partial T_b(t)}{\partial t} = Q_{\mathrm{sh}}(t) + Q_{v}(t) 
    + Q_{\mathrm{int}}(t) - Q_{\mathrm{out}}(t)
\end{eqnarray}
\setlength{\arraycolsep}{5pt}
\noindent where $Q_{\mathrm{sh}}(t)$ is the space heating load that we want to predict with the nominal model, $C_{\mathrm{th}}$ is the thermal mass of the building, $T_b(t)$ the lumped temperature of the building, $Q_v(t)$ is heat flow from ventilation air, $Q_{\mathrm{int}}(t)$ the energy from e.g. electrical equipment and residents, and $Q_{\mathrm{out}}(t)$ the temperature losses to ambient. The temperature losses to ambient can be approximated \cite{gustafsson_thermodynamic_2008} with a linear function of the outdoor temperature and the lumped building temperature
\begin{equation}
\label{eq:heat_transfer}
Q_{out}(t) = k_{ht}A_{ht}(T_b(t) - T_o(t)).
\end{equation}
If the supply temperature or substation only uses a feed forward based on the outdoor temperature, identification of the outdoor temperature dependency of the heat load is straight forward $Q_{\mathrm{sh}}(t) = \theta_1 T_o(t)$ under the assumption that the feed forward curve is reasonably linear. For the feedback case we can view the system as the block diagram shown in Figure \ref{fig:feedback}. 

\begin{figure}[ht!]
    \centering
    \begin{tikzpicture}[scale=0.425]

\draw (1,4) -- (4,4) -- (4,6) -- (1,6) -- (1,4);
\draw (5,4) -- (8,4) -- (8,6) -- (5,6) -- (5,4);
\draw (13,4) -- (16,4) -- (16,6) -- (13,6) -- (13,4);
\draw (5, 8.5) -- (8, 8.5) -- (8, 7.5) -- (5, 7.5) -- (5, 8.5);

\draw [->] (4,5) -- (5,5);
\draw [->] (8,5) -- (10.5,5);
\draw [->] (11.5,5) -- (13,5);
\draw [->] (16,5) -- (19,5);

\draw (11, 5) circle (0.5cm);
\draw (2.5, 2) circle (0.5cm);
\draw (3.5, 8) circle (0.5cm);

\draw (17, 5) -- (17, 2);
\draw [->] (17, 2) -- (3, 2);
\draw (2, 2) -- (0, 2);  
\draw (0, 2) -- (0, 5);
\draw [->] (0, 5) -- (1, 5);

\draw (17, 5) -- (17, 9);
\draw (17, 9) -- (3.5, 9);
\draw [->] (3.5, 9) -- (3.5, 8.5);

\draw [->] (4, 8) -- (5, 8);
\draw  (8, 8) -- (11, 8);
\draw [->] (11, 8) -- (11, 5.5);

\draw [->] (11, 3.5) -- (11, 4.5);
\draw [->] (2, 8) -- (3, 8);
\draw [->] (2.5, 0.5) -- (2.5, 1.5);

\node at (2.5, 5) {$C(s)$};
\node at (6.5, 5) {$G_s(s)$};
\node at (14.5, 5) {$G_b(s)$};
\node at (6.5, 8) {$-k_{th}A_{th}$};
\node at (8.75, 5.5) {$Q_{sh}$};
\node at (18, 5.5) {$T_b$};
\node at (2.5, 0) {$T_r(s)$};
\node at (1, 8) {$T_o(s)$};
\node at (11, 3) {$V(s)$};

\node at (11, 5) {$\Sigma$};
\node at (3.5, 8) {$\Sigma$};
\node at (2.5, 2) {$\Sigma$};
\node at (3.5, 2.5) {$-$};
\node at (2.5, 7.5) {$-$};
\node at (2.5, 10) {$\hspace{1cm}$};

\draw [dashed] (0.5, 6.5) -- (9.5, 6.5) -- (9.5, 1) -- (0.5, 1) -- (0.5, 6.5);  

\draw [dotted] (2.25, 7) -- (10, 7) -- (10, 3.75) -- (17.25, 3.75) -- (17.25, 9.5) -- (2.25, 9.5) -- (2.25, 7);
\end{tikzpicture}
    \caption{Block diagram of the feedback system.}
    \label{fig:feedback}
\end{figure}
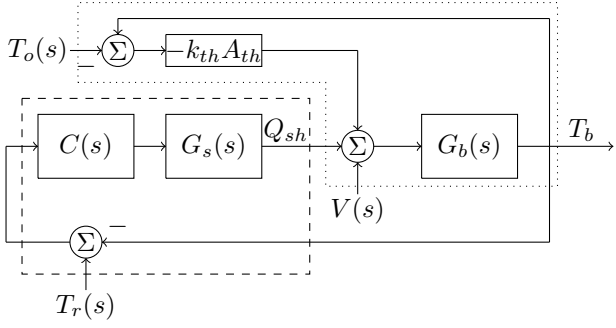{}

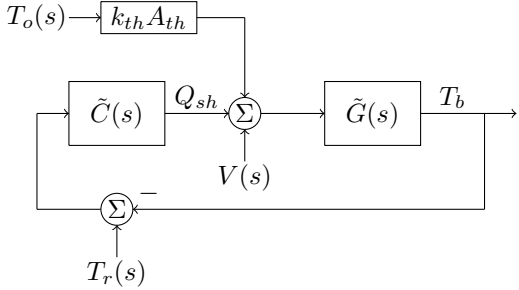
\begin{figure}[ht!]
    \centering
    \begin{tikzpicture}[scale=0.425]
\draw (1,4) -- (4,4) -- (4,6) -- (1,6) -- (1,4);
\draw (9,4) -- (12,4) -- (12,6) -- (9,6) -- (9,4);
\draw (2, 8.5) -- (5, 8.5) -- (5, 7.5) -- (2, 7.5) -- (2, 8.5);
\draw [->] (4, 5) -- (6, 5);
\draw [->] (7, 5) -- (9, 5);
\draw [->] (12,5) -- (15,5);
\draw (6.5, 5) circle (0.5cm);
\draw (2.5, 2) circle (0.5cm);
\draw (14, 5) -- (14, 2);
\draw [->] (14, 2) -- (3, 2);
\draw (2, 2) -- (0, 2);  
\draw (0, 2) -- (0, 5);
\draw [->] (0, 5) -- (1, 5);
\draw [->] (1, 8) -- (2, 8);
\draw  (5, 8) -- (6.5, 8);
\draw [->] (6.5, 8) -- (6.5, 5.5);

\draw [->] (6.5, 3.5) -- (6.5, 4.5);

\draw [->] (2.5, 0.5) -- (2.5, 1.5);

\node at (2.5, 5) {$\Tilde{C}(s)$};
\node at (10.5, 5) {$\Tilde{G}(s)$};
\node at (3.5, 8) {$k_{th}A_{th}$};

\node at (5, 5.5) {$Q_{sh}$};
\node at (13, 5.5) {$T_b$};
\node at (2.5, 0) {$T_r(s)$};
\node at (0, 8) {$T_o(s)$};
\node at (6.5, 3) {$V(s)$};

\node at (6.5, 5) {$\Sigma$};
\node at (2.5, 2) {$\Sigma$};
\node at (3.5, 2.5) {$-$};

\end{tikzpicture}
    \caption{Simplified representation of the feedback system.}
    \label{fig:feedback_simplified}
\end{figure}

If we first consider the blocks within the dotted line in Figure \ref{fig:feedback}, in Laplace form we have the building temperature
\begin{eqnarray}
    T_b(s) = (Q_{sh}(s) + V(s)\nonumber \\ 
    + k_{ht}A_{ht}(T_b(s) - T_o(s)))G_b(s)
\end{eqnarray}{}
\noindent that we can rearrange to get
\begin{eqnarray}
    T_b(s) = \Tilde{G}(s)(Q_{sh}(s) + V(s) + k_{ht}A_{ht}T_o(s))
\end{eqnarray}
\noindent where, by inserting (\ref{eq:energybalance}) in Laplace form, we have
\begin{eqnarray}
    \label{eq:g_tilde}
    \Tilde{G}(s) = {\frac{G_b(s)}{(1 + k_{ht}A_{ht} G_b(s))}} = \frac{1}{k_{ht}A_{ht} + C_{th}s}.
\end{eqnarray}{}
The substation dynamics are fast compared to the building dynamics, so we approximate the blocks within the dashed line in Figure \ref{fig:feedback} as $\Tilde{C}(s) = C(s)G_s(s)$. We have thus simplified the block diagram from Figure \ref{fig:feedback} to the form seen in Figure \ref{fig:feedback_simplified}.
{}

We can now look at $Q_{sh}$ that is our target quantity for the prediction, once again in Laplace form
\begin{eqnarray}
    Q_{sh}(s) = \Tilde{C}(s)\Tilde{G}(s)(T_r(s) - (Q_{sh}(s)\nonumber \\ 
    + k_{ht}A_{ht}T_o(s) + V(s))),
\end{eqnarray}
\noindent where rearranging the equation gives us three parts
\begin{eqnarray}
    \label{eq:q_split}
    Q_{sh}(s) = \underbrace{-\frac{\Tilde{C}(s)\Tilde{G}(s)}{1 + \Tilde{C}(s)\Tilde{G}(s)}V(s)}_{Q_{sh}^R(s)}\nonumber\\ 
    - \underbrace{\frac{k_{ht}A_{ht}\Tilde{C}(s)\Tilde{G}(s)}{1 + \Tilde{C}(s)\Tilde{G}(s)}}_{Q_{sh}^T(s)}T_o(s)
    + \frac{\Tilde{C}(s)}{1 + \Tilde{C}(s)\Tilde{G}(s)}T_r(s).
\end{eqnarray}{}
Here, $Q_{sh}^T(s)$ is the outdoor temperature dependent space heating load, that is target quantity for the nominal model. The set point of the indoor temperature $T_r(s)$ can be assumed to be constant in relation to the outdoor temperature $T_o(s)$, and the building temperature $T_b(s)$ can be assumed to be close to the reference temperature $T_r(s)$ so that the third term is relatively small. We then leave the prediction of $Q_{sh}^{R}$ as a time dependent residual part that is modeled by the latent variable model. The most common approach for feedback control of the indoor temperature is a PID controller, so that we get 
\begin{eqnarray}
    \label{eq:pid}
    \Tilde{C}(s) = K_p + \frac{K_i}{s} + K_d s,
\end{eqnarray}{}
\noindent where inserting (\ref{eq:pid}) and (\ref{eq:g_tilde}) into $Q_{sh}^T(s)$ from the equation above we get
\begin{eqnarray}
    Q_{sh}^T(s) = \nonumber\\
    \frac{-k_{ht}A_{ht}(K_i + K_p s + K_d s^2)}{K_i + (k_{ht}A_{ht} + K_p)s + (C_{th} + K_d)s^2}T_o(s)
\end{eqnarray}{}

\noindent that can be approximated by an ARX model of sufficiently high order. Further, exploiting the knowledge that there is generally no space heating when the outdoor temperature is high enough, we can substitute the outdoor temperature with $\Delta T(t) = \text{max}(T_c - T_o(t), 0)$ where $T_c$ is a user specified threshold temperature, to avoid the discontinuity. The nominal model can then be written as

\begin{equation}
\Hat{y}_{\mathrm{nom}}(t) = \Theta \varphi(t) = \Theta 
    \begin{bmatrix}
    1\\
    \Delta T(t)\\
    \Delta T(t-1)\\
    \vdots\\
    \Delta T(t-n_b)
    \end{bmatrix}. 
\end{equation}
Using $n_b=24$, that is 24h of time lags, has shown to give good predictions even during sharp outdoor temperature gradients for the building and substation used in the examples. Results show that including precipitation, wind speed or sun radiation has a negligible impact on the prediction accuracy of the example building. However, this is influenced by the control strategy in conjunction with factors such as where the building is located geographically, if it is shielded from wind, and the isolation of the building. Luleå is also located in the north of Sweden where the sun radiation during winter is low, whereas in the summer when the sun radiation is high there is no need for space heating.
\section{Latent variable model}
\label{sec:lavamodel}
The residual heat load that is predicted by the latent variable model is 
\begin{equation}
    y_{res}(t) = Q_{sh}^R(t) + Q_{tw}(t)   
\end{equation}{}
\noindent including the residual part from (\ref{eq:q_split}), and $Q_{tw}(t)$ that is the heat used for domestic hot water production. At the substation level it is concluded in \cite{gadd_heat_2013} that there exists four main heat load patterns based on the most commonly used control strategies for radiator and ventilation systems: continuous operation control, night set-back control, time clock operation control 5 days a week and time clock operation control 7 days a week. 

The hot tap water heat load patterns are highly dependent on the type of building, where a school is expected to have distinctively different hot tap water consumption patterns than a multi-dwelling apartment building. However, these patterns tend to show periodic behavior depending on time dependent covariates such as time of the day, day of the week, period of the year and weekend / not weekend, something that also holds for the time dependent control strategies above. These covariates are used as input $u_\gamma(t)$ to the latent variable model. 

The binary and continuous or periodic inputs to the latent variable model are denoted $u_{\gamma b}(t)$ and $u_{\gamma p}(t)$, with sizes $n_{\gamma b}$ and $n_{\gamma p}$ respectively. By using a Fourier expansion of $u_\gamma(t)$ as input to the latent variable model we can fit both periodic and continuous behavior from the covariates. Periodic and continuous signals are then approximated by the vector $\gamma_p(t)$ of orthogonal Fourier series basis expansions up to $M$ harmonics. Using the time of day expressed in hours $t_d(t)$, day of week $d_w(t)$, and week number $w_y(t)$ for the periodic or continuous part, and weekend/not weekend $wk(t)$ and summer/not summer $s(t)$ as a binary inputs, we get
\begin{equation}
    u_{\gamma p}(t) = \begin{bmatrix}t_d(t)\\d_w(t)\\w_y(t)\end{bmatrix} 
\quad \text{and} \quad
    u_{\gamma b}(t) = \begin{bmatrix}wk(t)\\s(t)\end{bmatrix}. 
\end{equation}
where the weekend signal includes official holidays, and May-August are considered as summer months. The basis expansion for each input of $u_{\gamma p, i}(t)$ and basis number $j$ can be written as
\begin{equation}
    b_{i,j}(t) =     
    \begin{bmatrix}
    \cos\left(\frac{j\pi u_{\gamma p, i} (t)}{2 \ell_i}\right)\\
    \sin\left(\frac{j\pi u_{\gamma p, i} (t)}{2 \ell_i}\right)\\
    \end{bmatrix}
\end{equation}
\noindent where $\ell_i$ is the boundary for each input. The $\gamma_p(t)$ vector contains the basis expansions of the periodic inputs
\begin{equation}
    \gamma_p(t) = 
    \begin{bmatrix}
        b_{1,1}(t)\\
        \vdots\\
        b_{1,M}(t)\\
        b_{2,1}(t)\\
        \vdots\\
        b_{n_{\gamma p},M}(t)
    \end{bmatrix}.
\end{equation} 
For the binary variables we have
\begin{eqnarray}
    \gamma_b(t) \! = \!
     \begin{bmatrix}1\\u_{\gamma b,1}(t)\\(1-u_{\gamma b, 1}(t))\end{bmatrix}   \otimes \hdots \otimes \begin{bmatrix}1\\u_{\gamma b,n\gamma b}(t)\\(1 - u_{\gamma b, n\gamma b}(t))\end{bmatrix} 
\end{eqnarray}
where $\otimes$ denotes the Kronecker product, and finally
\begin{eqnarray}
    \gamma(t) =  \left(\begin{bmatrix}
    \mathbf{0}_B&I_B 
    \end{bmatrix}  \gamma_b(t)\right) \otimes \gamma_p(t)
\end{eqnarray}
where $\mathbf{0}_B$ and $\mathbf{I}_B$ are $n_b \times 1$ and $n_b \times n_b$ respectively, with $n_{b}=((2+1)^{n_{\gamma b}}-1)$. The latent variable model will then have $n_\gamma = 2Mn_{\gamma p}((2+1)^{n_{\gamma b}}-1)$ variables, where in the results section we have used $M=8$ and thus $128$ variables in $Z$. For the case study including cross-terms of $\gamma(t)$ have not provided better prediction accuracy and have been left out for higher computational performance.

\section{Results}
\label{sec:results}
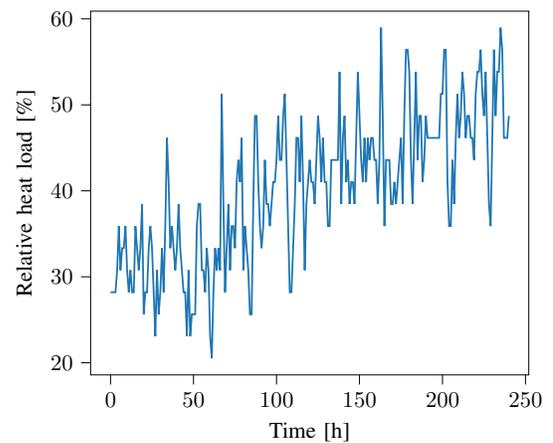
\begin{figure}[!ht]
    \centering
\begin{tikzpicture}[scale=0.85]

\definecolor{color0}{rgb}{0.12156862745098,0.466666666666667,0.705882352941177}

\begin{axis}[
tick align=outside,
tick pos=left,
x grid style={white!69.01960784313725!black},
xlabel={Time [h]},
xmin=-12, xmax=252,
xtick style={color=black},
y grid style={white!69.01960784313725!black},
ylabel={Relative heat load [\%]},
ymin=18.5897435897436, ymax=60.8974358974359,
ytick style={color=black}
]
\addplot [thick, color0]
table {%
0 28.2051282051282
1 28.2051282051282
2 28.2051282051282
3 28.2051282051282
4 30.7692307692308
5 35.8974358974359
6 30.7692307692308
7 33.3333333333333
8 33.3333333333333
9 35.8974358974359
10 30.7692307692308
11 28.2051282051282
12 30.7692307692308
13 28.2051282051282
14 28.2051282051282
15 35.8974358974359
16 33.3333333333333
17 30.7692307692308
18 33.3333333333333
19 38.4615384615385
20 25.6410256410256
21 28.2051282051282
22 28.2051282051282
23 33.3333333333333
24 35.8974358974359
25 33.3333333333333
26 28.2051282051282
27 23.0769230769231
28 30.7692307692308
29 25.6410256410256
30 28.2051282051282
31 33.3333333333333
32 28.2051282051282
33 35.8974358974359
34 46.1538461538462
35 41.025641025641
36 33.3333333333333
37 35.8974358974359
38 33.3333333333333
39 30.7692307692308
40 33.3333333333333
41 38.4615384615385
42 33.3333333333333
43 30.7692307692308
44 28.2051282051282
45 28.2051282051282
46 23.0769230769231
47 30.7692307692308
48 23.0769230769231
49 25.6410256410256
50 25.6410256410256
51 25.6410256410256
52 35.8974358974359
53 38.4615384615385
54 38.4615384615385
55 30.7692307692308
56 30.7692307692308
57 28.2051282051282
58 33.3333333333333
59 30.7692307692308
60 23.0769230769231
61 20.5128205128205
62 28.2051282051282
63 33.3333333333333
64 30.7692307692308
65 33.3333333333333
66 30.7692307692308
67 51.2820512820513
68 38.4615384615385
69 28.2051282051282
70 33.3333333333333
71 38.4615384615385
72 30.7692307692308
73 35.8974358974359
74 35.8974358974359
75 33.3333333333333
76 41.025641025641
77 43.5897435897436
78 41.025641025641
79 46.1538461538462
80 30.7692307692308
81 35.8974358974359
82 33.3333333333333
83 30.7692307692308
84 25.6410256410256
85 25.6410256410256
86 38.4615384615385
87 48.7179487179487
88 48.7179487179487
89 41.025641025641
90 35.8974358974359
91 33.3333333333333
92 35.8974358974359
93 43.5897435897436
94 38.4615384615385
95 38.4615384615385
96 35.8974358974359
97 38.4615384615385
98 41.025641025641
99 41.025641025641
100 43.5897435897436
101 48.7179487179487
102 43.5897435897436
103 43.5897435897436
104 48.7179487179487
105 51.2820512820513
106 43.5897435897436
107 35.8974358974359
108 28.2051282051282
109 28.2051282051282
110 33.3333333333333
111 38.4615384615385
112 46.1538461538462
113 46.1538461538462
114 41.025641025641
115 48.7179487179487
116 41.025641025641
117 30.7692307692308
118 38.4615384615385
119 41.025641025641
120 43.5897435897436
121 41.025641025641
122 41.025641025641
123 38.4615384615385
124 43.5897435897436
125 48.7179487179487
126 46.1538461538462
127 41.025641025641
128 46.1538461538462
129 41.025641025641
130 41.025641025641
131 35.8974358974359
132 35.8974358974359
133 43.5897435897436
134 43.5897435897436
135 43.5897435897436
136 43.5897435897436
137 43.5897435897436
138 53.8461538461538
139 38.4615384615385
140 46.1538461538462
141 48.7179487179487
142 41.025641025641
143 43.5897435897436
144 38.4615384615385
145 41.025641025641
146 41.025641025641
147 38.4615384615385
148 46.1538461538462
149 53.8461538461538
150 48.7179487179487
151 43.5897435897436
152 41.025641025641
153 46.1538461538462
154 41.025641025641
155 46.1538461538462
156 43.5897435897436
157 46.1538461538462
158 46.1538461538462
159 43.5897435897436
160 43.5897435897436
161 38.4615384615385
162 43.5897435897436
163 58.974358974359
164 48.7179487179487
165 35.8974358974359
166 43.5897435897436
167 43.5897435897436
168 43.5897435897436
169 38.4615384615385
170 38.4615384615385
171 41.025641025641
172 38.4615384615385
173 41.025641025641
174 43.5897435897436
175 46.1538461538462
176 38.4615384615385
177 48.7179487179487
178 56.4102564102564
179 56.4102564102564
180 53.8461538461538
181 43.5897435897436
182 38.4615384615385
183 46.1538461538462
184 53.8461538461538
185 43.5897435897436
186 48.7179487179487
187 48.7179487179487
188 41.025641025641
189 43.5897435897436
190 48.7179487179487
191 46.1538461538462
192 46.1538461538462
193 46.1538461538462
194 46.1538461538462
195 46.1538461538462
196 46.1538461538462
197 46.1538461538462
198 46.1538461538462
199 51.2820512820513
200 51.2820512820513
201 56.4102564102564
202 56.4102564102564
203 41.025641025641
204 35.8974358974359
205 35.8974358974359
206 43.5897435897436
207 38.4615384615385
208 46.1538461538462
209 51.2820512820513
210 46.1538461538462
211 48.7179487179487
212 53.8461538461538
213 51.2820512820513
214 46.1538461538462
215 48.7179487179487
216 48.7179487179487
217 46.1538461538462
218 46.1538461538462
219 43.5897435897436
220 51.2820512820513
221 53.8461538461538
222 53.8461538461538
223 56.4102564102564
224 51.2820512820513
225 48.7179487179487
226 53.8461538461538
227 46.1538461538462
228 38.4615384615385
229 35.8974358974359
230 46.1538461538462
231 56.4102564102564
232 48.7179487179487
233 53.8461538461538
234 53.8461538461538
235 58.974358974359
236 56.4102564102564
237 46.1538461538462
238 46.1538461538462
239 46.1538461538462
240 48.7179487179487
};
\end{axis}

\end{tikzpicture}
    \caption{Sample of heat load data series.}
    \label{fig:data_sample}
\end{figure}

Hourly heat load measurements were acquired from a single multi-dwelling building connected to the district heating network of Luleå, Sweden. The data was provided by Luleå Energi AB and anonymized in compliance with the General Data Protection Regulation (GDPR). Furthermore, weather data from the Swedish Metereological Institute (SMHI) and temporal data such as date and time was used. The heat load measurement has a dead band of 10kW causing a quantization effect on the signal, where the percentage relative heat load can be seen in Figure \ref{fig:data_sample}. 


The model has been trained with one year of measurement data, and is then continuously predicting 24h ahead out of sample for the following year, updating the model for every step ahead (walk-forward prediction). Each sample in the plot is the 24h ahead prediction.  Predictions use the \emph{actual} temperature ahead, where in real life weather predictions needs to be used. 

As a reference for the prediction accuracy a feed forward Artificial Neural Network (ANN) has been used. For a single hidden layer the ANN can be represented \cite{idowu_applied_2016} as
\begin{equation}
f(x) = \sum_{j=1}^{N}{w_j\psi_j}\left[\sum_{i=1}^{M}{w_{ij}x_i + w_{io}}\right] + w_{jo}
\end{equation}
\noindent where $M$ is the number of inputs, $N$ the number of hidden units and $\psi$ the transfer function for each hidden unit. The mean square error (MSE) has been used as cost function, and the tanh(x) as activation function. The neural network has been implemented with 1-3 hidden layers, that has previously been shown to produce good prediction results in \cite{idowu_applied_2016}. The results presented are from the best prediction, which in this case was using three layers. The ANN has been implemented using the MATLAB Deep Learning Toolbox. The same covariates as for the latent variable model have been used.

In order to evaluate the results, the relative root mean square error (rRMSE) of the data has been used as a performance metric. Since this is a metric commonly used for time series prediction the results are comparable to other methods, and indeed show that the estimates are in line with what has been previously reported. The reported rRMSE is for the whole validation data set, including the summer months with low heat load.
\begin{equation}
\text{rRMSE} = \frac{1}{\Bar{y}}\sqrt{\frac{1}{n}\sum_{i=1}^{n}{(y_i - \hat{y}_i)^2}}
\end{equation}
The resulting rRMSE\% can be seen in Table \ref{table:prediction_fit}, where the latent variable approach performs significantly better than the ANN method for this test case.

\begin{table}[!ht]
\centering 
\caption{Prediction, relative RMSE results}
\label{table:prediction_fit}
\begin{tabular}{c | c}
    \toprule
    LAVA Validation dataset &  18.2\%\\
    ANN Validation dataset &  28.8\%\\
    \bottomrule
\end{tabular}
\end{table}

\begin{figure}[!ht]
    \centering
    \input{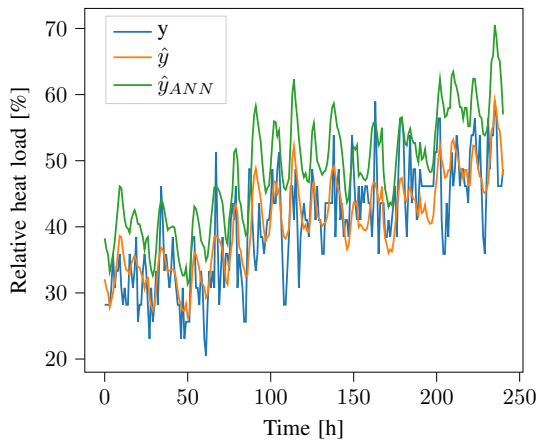}
    \caption{24h forecast and measured heat load for 10 winter days.}
    \label{fig:forecast}
\end{figure}

\begin{figure}[!ht]
    \centering
    \input{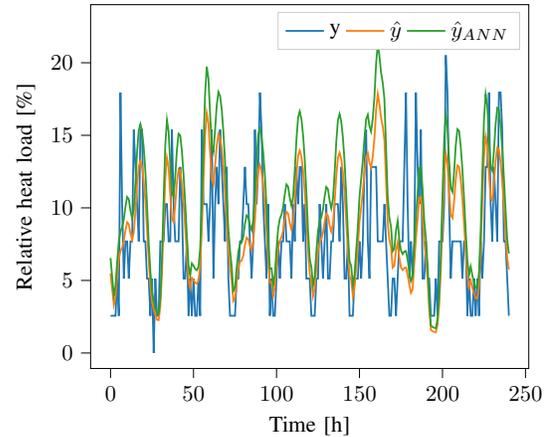}
    \caption{24h forecast and measured heat load for 10 summer days.}
    \label{fig:forecast_summer}
\end{figure}

Two prediction plots are provided, one time period of 10 days with a relatively sharp temperature gradient, and one for a summer month without any space heating load. 
Prediction results can be seen in Figure \ref{fig:forecast} and \ref{fig:forecast_summer}, with the results from the ANN prediction as a reference. Notably, some very sharp morning peaks in load occur during some weekdays, but not other weekdays, nor the same weekdays the week after. Such irregularities in consumer behaviour can not be predicted from cyclic variables like the time of the day, and would need additional covariates for prediction, if they are at all predictable.

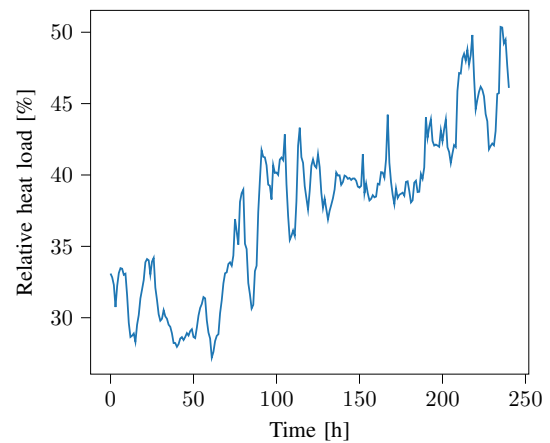
\begin{figure}[!ht]
    \centering
\begin{tikzpicture}[scale=0.85]

\definecolor{color0}{rgb}{0.12156862745098,0.466666666666667,0.705882352941177}

\begin{axis}[
tick align=outside,
tick pos=left,
x grid style={white!69.01960784313725!black},
xlabel={Time [h]},
xmin=-12, xmax=252,
xtick style={color=black},
y grid style={white!69.01960784313725!black},
ylabel={Relative heat load [\%]},
ymin=26.0483159303115, ymax=51.5311557426693,
ytick style={color=black}
]
\addplot [thick, color0]
table {%
0 33.0940473916123
1 32.8311618849671
2 32.2993749117549
3 30.7473813026361
4 32.2251888471678
5 33.1317399136903
6 33.473361851424
7 33.4330579516824
8 33.0055142863947
9 33.0828789378708
10 31.5579308443456
11 29.6361414417969
12 28.6487600129887
13 28.7412472747226
14 28.8949276352232
15 28.3146159469678
16 29.5048833683074
17 30.183996793456
18 31.3555172106718
19 31.9826500555323
20 32.6381624857092
21 33.8868440945825
22 34.1120360882
23 34.0223513298409
24 33.0310807072957
25 33.9338493697897
26 34.1906499450348
27 32.1093984037595
28 31.2436483425071
29 30.2429808591647
30 29.7905281731127
31 29.9295287816191
32 30.5246958879981
33 30.0959934664377
34 29.92958064486
35 29.5121081676911
36 29.364962038068
37 28.9076436560467
38 28.2265246284752
39 28.2250468607628
40 27.9613554714948
41 28.1519642299527
42 28.5442078922508
43 28.6436206487996
44 28.4404520679088
45 28.6519118987798
46 28.9212493463732
47 28.7533079696586
48 29.0614840326095
49 29.195049411472
50 28.6497179878172
51 28.5724737291593
52 29.2988792689372
53 30.1493577194104
54 30.684411408002
55 30.9633481848931
56 31.4474567716214
57 31.358326168808
58 29.8295895718557
59 28.9645926733418
60 28.5443793804493
61 27.2066268308732
62 27.6082390431764
63 28.3649836170704
64 28.7320509288742
65 28.8460691239178
66 30.3116969959737
67 31.2155993008818
68 32.405611959629
69 33.1121102375066
70 33.1775504927444
71 33.761447288389
72 33.8817546768273
73 33.6473793303887
74 34.3513846118585
75 36.9126460333898
76 36.0443485558936
77 35.1214569043447
78 38.1190653721849
79 38.726587387884
80 38.9733372868784
81 35.1826390412836
82 34.8118582458104
83 32.4289993955944
84 31.6317412244386
85 30.6491654512871
86 30.9047461825386
87 33.2640398829092
88 33.6412107418151
89 37.4162561740146
90 39.6514695742512
91 41.7290292313095
92 41.2739508547591
93 41.2340657709529
94 40.6814068814009
95 39.3253457602972
96 39.2420615901895
97 38.2824926753966
98 40.692393109293
99 40.1439190796291
100 40.1738061860236
101 40.0098059254788
102 41.0790892149366
103 41.2226851685097
104 41.0446349926962
105 42.844053000838
106 39.3679563855181
107 37.1276790758472
108 35.4514324651774
109 35.7375038133225
110 36.1101903929607
111 35.7400788469466
112 38.1457347317116
113 42.0501206102211
114 43.312586921954
115 41.245446547806
116 40.846578147636
117 39.2233709415931
118 38.3864452719729
119 37.5756816190692
120 38.8837541482953
121 40.6354250701198
122 41.0939626609607
123 40.6424728481591
124 40.5079494903286
125 41.4554627311677
126 40.5493594639706
127 38.7600062123082
128 37.6915039966571
129 38.3387660466199
130 37.6102043106238
131 36.885432158715
132 37.4908394075364
133 37.9136943173673
134 38.3678358976457
135 38.9823560039218
136 40.1739223636162
137 39.9599556895715
138 39.9751056381598
139 39.3127334768773
140 39.5048515968029
141 39.9571064823722
142 39.8901145186232
143 39.7279570430226
144 39.7920964866093
145 39.6505410866276
146 39.7562165428566
147 39.7533900701018
148 39.5955006459266
149 39.1871468320843
150 39.1116615366333
151 39.2356304359794
152 41.4590673342382
153 38.7240765675499
154 39.4009572176966
155 38.706455287233
156 38.2152359328839
157 38.3267866227577
158 38.5832115711985
159 38.4340308461128
160 38.489084405556
161 39.3825946265528
162 39.3305275403093
163 40.1978840804682
164 40.1751731617428
165 39.8202755681837
166 41.104920799777
167 44.2236846966815
168 40.8611971279554
169 39.5122378064963
170 38.6044145489503
171 37.9511612766635
172 39.0588855738611
173 38.3919135835515
174 38.6147005692972
175 38.6830970806917
176 38.7602578229996
177 38.5552496828196
178 39.5149805358976
179 39.5572478607455
180 38.8676296939154
181 38.0797248297622
182 38.2279767478347
183 39.4625128752143
184 39.5972504436592
185 38.8031975991208
186 38.8255883404279
187 40.073906037505
188 39.74330818553
189 40.5244794324757
190 44.0322291833579
191 42.6164158542225
192 43.3582982938708
193 43.8941483543458
194 42.3811700519998
195 42.0608344978408
196 42.1305620817612
197 42.0624027695881
198 41.976199112732
199 43.1234967238625
200 42.3740143702859
201 43.2050446566487
202 43.8841583330649
203 41.9444989174195
204 41.6100547712788
205 40.8357252994612
206 41.5021631752664
207 42.0948512015505
208 41.9576839031928
209 45.8852866480665
210 47.1433369472307
211 47.1059747121312
212 48.1625614329124
213 48.4923936427851
214 48.0052091018974
215 48.7741978164021
216 47.717261427897
217 48.3922999175171
218 49.8082839101847
219 46.9022768925671
220 44.6034457790066
221 45.2704768759262
222 45.8244359099584
223 46.1827058699561
224 45.970980664332
225 45.5108479150508
226 44.2668862231404
227 43.767897357289
228 41.8042646893529
229 42.0300989408259
230 42.2052394179421
231 42.0768257982299
232 43.0750407781459
233 45.6739701581694
234 45.7335158453271
235 50.3728448421076
236 50.3311584602178
237 49.2209235024624
238 49.4794668947791
239 47.6753718524836
240 46.1010193000741
};
\end{axis}

\end{tikzpicture}
    \caption{Prediction contribution from the nominal model.}
    \label{fig:Theta_phi}
\end{figure}

The output from the nominal model $\hat{y}_{nom}(t) =\Theta\varphi(t)$ is seen in Figure \ref{fig:Theta_phi}. As expected the nominal model follows the $\Delta T$ temperature closely, but also has a dynamic component. 

\begin{figure}[!ht]
    \centering
\begin{tikzpicture}[scale=0.85]
every node/.style={inner sep=0,outer sep=0}

\definecolor{color0}{rgb}{0.12156862745098,0.466666666666667,0.705882352941177}

\begin{axis}[
tick align=outside,
tick pos=left,
x grid style={white!69.01960784313725!black},
xlabel={Time [h]},
xmin=-12, xmax=252,
xtick style={color=black},
y grid style={white!69.01960784313725!black},
ylabel={Relative heat load [\%]},
ymin=-4.2336477331019, ymax=9.91260210365702,
ytick style={color=black}
]
\addplot [thick, color0]
table {%
0 -1.6093213951605
1 -2.45257391990646
2 -2.58694273171846
3 -3.12656217222206
4 -3.59063637688559
5 -3.40158045868533
6 -2.16286375766691
7 0.0862763870970818
8 2.80134986741186
9 5.14769871351494
10 6.42303503334813
11 6.44128468503985
12 5.61988342440447
13 4.69751553716079
14 4.26551359134458
15 4.42705912636746
16 4.79978936620235
17 4.82751315188425
18 4.1600915797279
19 2.84606521058253
20 1.24803908069801
21 -0.210531514858993
22 -1.27424976516434
23 -1.93091925856613
24 -2.33173947371579
25 -1.91823942859493
26 -2.05260824040693
27 -2.59222768091053
28 -3.05630188557406
29 -2.8672459673738
30 -1.62852926635539
31 0.620610878408608
32 3.33568435872339
33 5.68203320482647
34 6.95736952465965
35 6.97561917635137
36 6.15421791571599
37 5.23185002847231
38 4.7998480826561
39 4.96139361767898
40 5.33412385751387
41 5.36184764319577
42 4.69442607103942
43 3.38039970189405
44 1.78237357200954
45 0.323802976452533
46 -0.739915273852817
47 -1.3965847672546
48 -1.79740498240426
49 -1.11718469236782
50 -2.56287519250084
51 -1.97941903799029
52 0.617498844941037
53 3.64025543062974
54 5.25109288336533
55 4.83900665183988
56 3.32807481673796
57 2.14902828048499
58 1.95090578083502
59 2.27015874175203
60 2.33914424004825
61 2.08182035614948
62 2.2340295393446
63 3.49518244529115
64 5.64058103138405
65 7.55472455544519
66 8.17882603060229
67 7.41446850628266
68 6.07275983244476
69 4.94985503704037
70 4.01566382944217
71 2.57981129355993
72 0.264514544184443
73 -1.82034188817292
74 -3.26603238830594
75 -2.68257623379539
76 -0.0856583508640652
77 2.93709823482464
78 4.54793568756023
79 4.13584945603478
80 2.62491762093286
81 1.44587108467988
82 1.24774858502992
83 1.56700154594693
84 1.63598704424315
85 1.37866316034438
86 1.5308723435395
87 2.79202524948605
88 4.93742383557895
89 6.85156735964009
90 7.47566883479719
91 6.71131131047756
92 5.36960263663966
93 4.24669784123526
94 3.31250663363707
95 1.87665409775483
96 -0.43864265162066
97 -0.929040660682908
98 -2.37473116081593
99 -1.79127500630538
100 0.805642876625948
101 3.82839946231465
102 5.43923691505024
103 5.02715068352479
104 3.51621884842287
105 2.3371723121699
106 2.13904981251993
107 2.45830277343694
108 2.52728827173316
109 2.26996438783439
110 2.42217357102951
111 3.68332647697606
112 5.82872506306896
113 7.7428685871301
114 8.3669700622872
115 7.60261253796757
116 6.26090386412967
117 5.13799906872528
118 4.20380786112709
119 2.76795532524484
120 0.452658575869354
121 -0.62696116233927
122 -2.07265166247229
123 -1.48919550796174
124 1.10772237496959
125 4.13047896065829
126 5.74131641339388
127 5.32923018186843
128 3.81829834676651
129 2.63925181051354
130 2.44112931086357
131 2.76038227178058
132 2.8293677700768
133 2.57204388617803
134 2.72425306937315
135 3.9854059753197
136 6.1308045614126
137 8.04494808547374
138 8.66904956063084
139 7.90469203631121
140 6.56298336247331
141 5.44007856706892
142 4.50588735947072
143 3.07003482358848
144 0.754738074212992
145 -1.93994918075706
146 -3.38563968089008
147 -2.80218352637953
148 -0.2052656434482
149 2.81749094224051
150 4.42832839497609
151 4.01624216345064
152 2.50531032834873
153 1.32626379209575
154 1.12814129244578
155 1.44739425336279
156 1.51637975165901
157 1.25905586776025
158 1.41126505095537
159 2.67241795690191
160 4.81781654299481
161 6.73196006705595
162 7.35606154221305
163 6.59170401789342
164 5.24999534405552
165 4.12709054865113
166 3.19289934105294
167 1.75704680517069
168 -0.558249944204795
169 -1.18445028908484
170 -1.31881910089684
171 -1.85843854140043
172 -2.32251274606397
173 -2.13345682786371
174 -0.894740126845294
175 1.3544000179187
176 4.06947349823348
177 6.41582234433656
178 7.69115866416975
179 7.70940831586147
180 6.88800705522609
181 5.9656391679824
182 5.5336372221662
183 5.69518275718907
184 6.06791299702397
185 6.09563678270587
186 5.42821521054952
187 4.11418884140415
188 2.51616271151963
189 1.05759211596263
190 -0.0061261343427225
191 -0.662795627744506
192 -1.06361584289417
193 -0.650115797773311
194 -0.784484609585311
195 -1.32410405008891
196 -1.78817825475244
197 -1.59912233655218
198 -0.360405635533768
199 1.88873450923023
200 4.60380798954501
201 6.95015683564809
202 8.22549315548127
203 8.24374280717299
204 7.42234154653761
205 6.49997365929393
206 6.06797171347772
207 6.2295172485006
208 6.60224748833549
209 6.62997127401739
210 5.96254970186104
211 4.64852333271567
212 3.05049720283116
213 1.59192660727415
214 0.528208356968803
215 -0.12846113643298
216 -0.529281351582645
217 -0.0264199755294028
218 -1.47211047566243
219 -0.888654321151875
220 1.70826356177945
221 4.73102014746816
222 6.34185760020374
223 5.92977136867829
224 4.41883953357638
225 3.2397929973234
226 3.04167049767344
227 3.36092345859045
228 3.42990895688667
229 3.1725850729879
230 3.32479425618302
231 4.58594716212957
232 6.73134574822247
233 8.64548927228361
234 9.2695907474407
235 8.50523322312108
236 7.16352454928318
237 6.04061975387878
238 5.10642854628059
239 3.67057601039835
240 1.35527926102286
};
\end{axis}

\end{tikzpicture}
    \caption{Prediction contribution from the latent variable model, showing heat load patterns for both weekend and workday.}
    \label{fig:Z_gamma}
\end{figure}
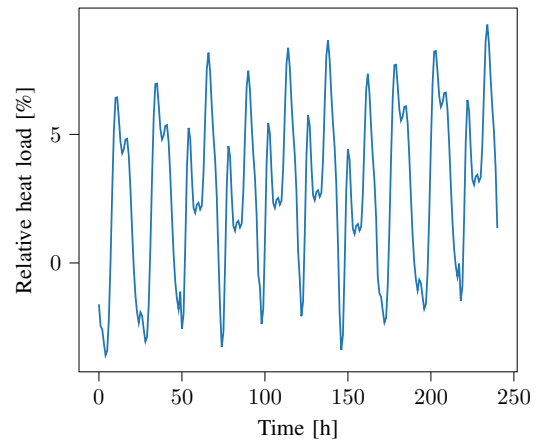

The output of the latent variable model $\hat{y}_{res}(t) = Z\gamma(t)$ that can be seen in Figure \ref{fig:Z_gamma} shows that the models have adapted to the diurnal pattern of the heat load, with distinct different patterns for workdays and the weekend. The resulting $Z$ in this case has 70 non-zero parameters.

\begin{figure}[!ht]
    \centering
\begin{tikzpicture}[scale=0.85]

\definecolor{color0}{rgb}{0.12156862745098,0.466666666666667,0.705882352941177}
\definecolor{color1}{rgb}{1,0.498039215686275,0.0549019607843137}
\definecolor{color2}{rgb}{0.172549019607843,0.627450980392157,0.172549019607843}

\begin{axis}[
legend cell align={left},
legend entries={{M=2},{M=4},{M=8}},
legend style={at={(0.03,0.97)}, anchor=north west, legend  columns =-1, draw=white!80.0!black},
tick align=outside,
tick pos=left,
title={\hspace{1cm}},
x grid style={white!69.01960784313725!black},
xlabel={Time [h]},
xmin=-3.6, xmax=75.6,
y grid style={white!69.01960784313725!black},
ylabel={Relative heat load [\%]},
ymin=-4, ymax=12 
]
\addlegendimage{no markers, color0}
\addlegendimage{no markers, color1}
\addlegendimage{no markers, color2}
\addplot [thick, color0]
table [row sep=\\]{%
0	1.6343839670631 \\
1	-0.351645570546321 \\
2	0.501381304983713 \\
3	1.19951759158867 \\
4	1.72430183734742 \\
5	2.22685197272945 \\
6	2.61688119463002 \\
7	2.99909369931497 \\
8	3.36867001968636 \\
9	3.70597479998549 \\
10	4.06294569638071 \\
11	4.41368484593771 \\
12	4.73968649023617 \\
13	5.19318819421177 \\
14	5.58908076054503 \\
15	5.96679248255761 \\
16	6.24576540238429 \\
17	6.38208390503708 \\
18	6.22611356794612 \\
19	5.89101131972238 \\
20	5.35351121596978 \\
21	4.65823338409464 \\
22	3.52658701653381 \\
23	2.21011857902096 \\
24	0.687477458290505 \\
25	-1.38846755664082 \\
26	-0.503542401712952 \\
27	0.271467919146657 \\
28	0.999843100344215 \\
29	1.42358478566241 \\
30	1.85498866345008 \\
31	2.2801175925821 \\
32	2.57009674831945 \\
33	2.98796343011074 \\
34	3.26468379116867 \\
35	3.53347253423772 \\
36	3.86186877666176 \\
37	4.2716587978008 \\
38	4.69803033444159 \\
39	5.15377030006976 \\
40	5.43233626733998 \\
41	5.53626763974129 \\
42	5.5304270134737 \\
43	5.23456165575486 \\
44	4.69387471195679 \\
45	3.80318542858822 \\
46	2.7123929951047 \\
47	1.48270454582495 \\
48	-0.118450378696264 \\
49	-2.24726333558546 \\
50	-2.56892586152236 \\
51	-2.14084985582393 \\
52	-1.41274501689669 \\
53	-0.402729539632888 \\
54	0.799836894029654 \\
55	2.17544907434602 \\
56	3.47527250205414 \\
57	4.72577450353627 \\
58	5.92930257484327 \\
59	6.9601299582083 \\
60	7.64453681758275 \\
61	8.08504861734167 \\
62	8.20527048599275 \\
63	7.97440125977516 \\
64	7.35752570094522 \\
65	6.57049648341198 \\
66	5.47037842456889 \\
67	4.36085226664819 \\
68	3.26358720858486 \\
69	2.08663274379672 \\
70	1.05447475262605 \\
71	0.207878337285475 \\
72	-0.425295826550197 \\
};
\addplot [thick, color1]
table [row sep=\\]{%
0	0.42603303336959 \\
1	-0.864336205049048 \\
2	-0.566459069090656 \\
3	0.530167281703099 \\
4	2.09019728862854 \\
5	3.71616647615953 \\
6	4.98805076678991 \\
7	5.63603119429263 \\
8	5.58553819767134 \\
9	4.95772378413243 \\
10	4.05343097541596 \\
11	3.25326423678381 \\
12	2.87922374418182 \\
13	3.20002572961774 \\
14	4.18810134087738 \\
15	5.66473581743126 \\
16	7.26316275100322 \\
17	8.55969286908283 \\
18	9.15582984325436 \\
19	8.85892316943002 \\
20	7.66881578871083 \\
21	5.80458344431085 \\
22	3.55136112452468 \\
23	1.47816316532254 \\
24	-0.00448602349371493 \\
25	-1.50463534569222 \\
26	-1.21019815721999 \\
27	-0.0954152820380895 \\
28	1.52577988622472 \\
29	3.12399571244404 \\
30	4.40365961901659 \\
31	5.07010455742084 \\
32	4.98978934681382 \\
33	4.3861161605569 \\
34	3.44137108112516 \\
35	2.57985640988748 \\
36	2.17447461575242 \\
37	2.46656551498837 \\
38	3.45278136107233 \\
39	4.9733005771505 \\
40	6.57044629888585 \\
41	7.8545908236975 \\
42	8.5001206323017 \\
43	8.20883534300407 \\
44	7.00590408363755 \\
45	5.03877989922333 \\
46	2.79034582312591 \\
47	0.781060005481071 \\
48	-0.746750621417028 \\
49	-0.710865770708365 \\
50	-2.88704981693677 \\
51	-3.38275917208047 \\
52	-2.82825321365895 \\
53	-1.31635528911218 \\
54	0.782670664967251 \\
55	3.16247149144217 \\
56	5.13009786334556 \\
57	6.59099516326265 \\
58	7.56006449990605 \\
59	7.96451629670251 \\
60	7.78417706658499 \\
61	7.48613694093107 \\
62	7.19299318883554 \\
63	6.99538544674031 \\
64	6.81880262398091 \\
65	6.81069796411882 \\
66	6.47865406010044 \\
67	5.9864474227337 \\
68	5.12027402551434 \\
69	3.62949667272283 \\
70	1.88796027706052 \\
71	0.138252391859467 \\
72	-1.3137922302937 \\
};
\addplot [thick, color2]
table [row sep=\\]{%
0	0.885679464716029 \\
1	0.422743097465757 \\
2	-1.24023921580178 \\
3	-0.437247579014686 \\
4	2.12740537308536 \\
5	4.78902874089152 \\
6	6.00232787011528 \\
7	5.7347219210302 \\
8	4.93195551135388 \\
9	4.45849571733244 \\
10	4.3947428778935 \\
11	4.16523760748709 \\
12	3.4598889389561 \\
13	2.93821974652316 \\
14	3.37826915741316 \\
15	5.17389938737285 \\
16	7.59982822072802 \\
17	9.39058554748836 \\
18	9.64012778602246 \\
19	8.56107301060548 \\
20	6.90815764471599 \\
21	5.27895540988797 \\
22	3.56313379505054 \\
23	1.74796467142767 \\
24	0.0383836843461552 \\
25	-0.695640047500677 \\
26	-2.33307506030402 \\
27	-1.47665288329562 \\
28	1.24836849465206 \\
29	3.86311056305304 \\
30	5.11570276750264 \\
31	4.87503284731328 \\
32	4.01599020058224 \\
33	3.59883159073811 \\
34	3.49984031171174 \\
35	3.26309953614772 \\
36	2.62106395302658 \\
37	2.08073287897233 \\
38	2.52584350445782 \\
39	4.33936958992793 \\
40	6.76469400711187 \\
41	8.56446931943725 \\
42	8.89155616040638 \\
43	7.84901594691793 \\
44	6.20246915960443 \\
45	4.50620164586864 \\
46	2.82358607048358 \\
47	1.0335763384817 \\
48	-0.70317614932913 \\
49	-0.554355185474866 \\
50	-1.01251907061566 \\
51	-1.35180992539672 \\
52	-1.52616487097096 \\
53	-1.23287896612617 \\
54	0.0580708368188285 \\
55	2.605319580737 \\
56	5.93898359858659 \\
57	8.85060593620978 \\
58	10.1344968943859 \\
59	9.46069798204375 \\
60	7.65220092115803 \\
61	6.1044495846117 \\
62	5.82885848671444 \\
63	6.78603870174595 \\
64	8.07224133169012 \\
65	8.67825631951844 \\
66	8.17039650406743 \\
67	6.76936101646888 \\
68	4.97607934994476 \\
69	3.15498037974659 \\
70	1.49437470561626 \\
71	0.224743473827021 \\
72	-0.284154199039914 \\
};
\end{axis}

\end{tikzpicture}
    \caption{Output from the latent variable model using different number of harmonics.}
    \label{fig:Z_gamma_comparison_M}
\end{figure}
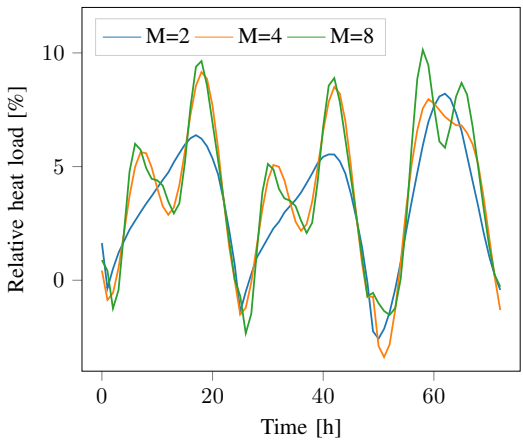

A comparison between the outputs from $\hat{y}_{res}(t) =Z\gamma(t)$ using different number of harmonics $M$ for the latent variable model can be seen in Figure \ref{fig:Z_gamma_comparison_M}, with the corresponding rRMSE metrics for the validation data set listed in Table \ref{table:M_fit}. Since more harmonics renders a larger $Z$ matrix this comes with a computational performance penalty.

\begin{table}[!ht]
\centering 
\caption{Prediction relative RMSE results}
\label{table:M_fit}
\begin{tabular}{c | c}
    \toprule
    M=2 & 19.0\%\\
    M=4 & 18.4\%\\
    M=8 & 18.2\%\\
    M=12 & 18.49\%\\
    \bottomrule
\end{tabular}
\end{table}

Over the year the largest deviations from the actual load are typically found around certain holidays, suggesting that the workday-weekend split might not be fine grained enough for the most accurate predictions. 



\section{Conclusions and future research}
In this article, it is shown how a hybrid approach using a nominal model for outdoor temperature dependent heat load together with a latent variable model for the residual heat load, can be used for prediction of the total heat load for a multi-dwelling building. The model structure is suitable both for offline simulation and in an online setup where the models are continuously updated. The implementation of the parameter estimation is recursive and hyper-parameter free, allowing for an easy parameterization of the model. The algorithm produces parsimonious models that can be efficiently simulated on a standard computer.

In order to simulate a city scale district heating network, with use cases such as the addition of new city quarters to the grid, further research on the classification of different types of buildings is needed. The proposed model structure allows for separate categorization of the nominal and latent variable model structures.

The model structure allows for a straight-forward aggregation of the distributed heat loads, whereas aggregation of consumers for a specific point in the network requires accounting for the distribution of the thermal heat. 

For a city scale simulation, different models, such as models with more detailed dynamics, consumer dependency on other covariates, and consumers where the heat load cannot be predicted from acquirable signals, need to be integrated in the same simulation. Accordingly, a generic consumer heat load framework needs to be considered, where the proposed model is a piece of a larger puzzle.
\bibliography{references}

\end{document}